\documentclass[journal]{rmaa}

\usepackage{paralist}
\usepackage{wrapfig}
\usepackage{multicol}
\usepackage{multirow}

\usepackage{psfrag,color}

\usepackage[latin1]{inputenc}

\usepackage{amsmath}
\usepackage{diagbox}

\title{Estimating the propagation of a uniformly \\accelerated jet} 

\author{
  J.I. Castorena,\altaffilmark{1} 
  A.C. Raga,\altaffilmark{1}
  A. Esquivel,\altaffilmark{1}
  A. Rodr\'iguez-Gonz\'alez\altaffilmark{1}
  L. Hern\'andez-Mart\'inez,\altaffilmark{2}
  \\ J. Cant\'o,\altaffilmark{3}
  F. Clever.\altaffilmark{3}}

\altaffiltext{1}{Instituto de Ciencias Nucleares, UNAM, M\'exico.}
\altaffiltext{2}{Facultad de Ciencias, UNAM, M\'exico}
\altaffiltext{3}{Instituto de Astronom\'ia, UNAM, M\'exico.}

\shortauthor{Castorena et al.}
\shorttitle{Accelerated jet}

\fulladdresses{
\item J.I. Castorena: Instituto de Ciencias Nucleares, Universidad Nacional Aut\'onoma de M\'exico, Apdo. Postal 70-543, 04510, CDMX, M\'exico.(jicastorena@correo.nucleares.unam.mx).
\item   A.C. Raga, A. Rodr\'iguez-Gonz\'alez, A. Esquivel: Instituto de Ciencias Nucleares, Universidad Nacional Aut\'onoma de M\'exico, Apdo. Postal 70-543, 04510, CDMX, M\'exico (raga, ary, esquivel@nucleares.unam.mx).
\item L. Hern\'andez-Mart\'inez: Facultad de Ciencias, Universidad Nacional Aut\'onoma de M\'exico, 
Av. Universidad 3000, Circuito Exterior S$/$N, Delegaci\'on Coyoac\'an, C.P. 04510, Ciudad Universitaria, D.F., M\'exico (lilihe@ciencias.unam.mx)
\item J. Cant\'o and F. Clever: Instituto de Astronom\'ia, Universidad Nacional Aut\'onoma de M\'exico, Apdo. Postal 70-264, 04510, CDMX, M\'exico (mclever@astro.unam.mx).}

\listofauthors{J.I. Castorena,  A.C. Raga, , A. Esquivel, L., A. Rodr\'iguez-Gonz\'alez, Hern\'andez-Mart\'inez, F. Clever, J. Cant\'o,}
\indexauthor{Castorena, J.I.}
\indexauthor{Raga, A.C.}
\indexauthor{Rodr\'iguez-Gonz\'alez, A.}
\indexauthor{Esquivel, A.}
\indexauthor{Hern\'andez-Mart\'inez, L.}
\indexauthor{Clever, F.}
\indexauthor{Cant\'o, J.}

\abstract{We study the problem of a Herbig-Haro jet with a uniformly accelerating ejection velocity, travelling into a uniform environment. For the ejection density we consider two cases: a time-independent density, and a time-independent mass loss rate. For these two cases, we obtain analytic solutions for the motion of the jet head using a ram-pressure balance and a center of mass equation of motion. We also compute axisymmetric numerical simulations of the same flow, and compare the time-dependent positions of the leading working surface shocks with the predictions of the two analytic models. We find that if the jet is over-dense and over-pressured (with respect to the environment) during its evolution, a good agreement is obtained with the analytic models, with the flow initially following the center of mass analytic solution, and (for the constant ejection
  density case) at later times approaching the ram-pressure balance solution.}

\resumen{Estudiamos el problema de un jet Herbig-Haro que viaja en un medio ambiente homog\'eneo con una velocidad de eyecci\'on uniformente acelerada. Consideramos dos casos para la densidad de eyecci\'on: una densidad constante, y una tasa de p\'erdida de masa constante. En ambos casos utilizamos tanto una ecuaci\'on de equilibrio de presiones hidrodin\'amicas como una de posici\'on de centro de masa para obtener soluciones anal\'iticas que describan la propagaci\'on de la cabeza del jet. Tambi\'en realizamos simulaciones num\'ericas de este flujo, y comparamos la dependencia temporal de la posici\'on de los choques que conforman la superficie de trabajo con las predicciones de los modelos anal\'iticos. Encontramos que si la densidad y la presi\'on del jet son mayores que los correspondientes valores ambientales durante toda la evoluci\'on, entonces una buena coincidencia con los modelos anal\'\i ticos es obtenida: en un inicio el flujo sigue la soluci\'on anal\'\i tica de centro de masa y en tiempos posteriores (para
  el caso de densidad de eyecci\'on constante) se acerca a la soluci\'on de balance de presiones hidrodin\'amicas.}

\addkeyword{ }
\addkeyword{ISM: Jets and outflows}
\addkeyword{HH objects}
\addkeyword{ }

\begin{document}
\maketitle

\section{Introduction}
\label{sec:intro}

Jets from young stars and their associated Herbig-Haro objects (HH)
are now the most studied and best understood of all astrophysical jets.
HH jets are collimated bipolar ejections from low, intermediate and
sometimes high mass protostars
or young stars (see, e.g., the review of Frank et al. 2014). The interaction between HH jets and the surrounding environment generates a main working surface known as the ``head''. This structure (formed by a bow shock/jet shock pair) is a clear sign of the ``turning on'' of the jet, which tipically occurred at times of $\sim 10^3\to 10^4$~yr ago in observed HH jets.

This type of jets also have a structure of emitting knots (some of them resembling the
jet head, and others forming aligned chains of more compact emission peaks)
in the region between the outflow source and the jet head
(see, e.g., Reipurth \& Heathcote 1990 and the review of Reipurth \&
Bally 2001). Several possibilities
of how to model these knots have been explored (see, e.g., Raga \& Kofman, 1992, and Micono et al. 1998),
but presently the favoured explanation is that they are the result of a time-variability
in the ejection, which leads to the formation of ``internal working surfaces''
within the jet beam (see, e.g., Raga et al 1990). Some of the ``classical HH objects''
of the catalogue of \citet{H74} are associated with either jet heads or knots.

A problem that has received little attention is the effect
of a ``slow turning on'' of the outflow on the jet head. As far as we
are aware, the only simulations studying the dynamics of the head of
a ``slow turning on'' jet (as opposed to a jet which is instantaneously
``switched on'' at full velocity) are the ones of \citet{LETAL02}.
These authors focus their study on the survival of H$_2$
molecules in the resulting, accelerating
jet head,  presenting 1D hydrodynamic simulations including an H$_2$
formation/destruction chemical network.

Of course, there are a number of numerical studies of the ejection of
MHD jets from star+accretion disk systems (e.g., Ouyed et al. 2003;
Hiromitsu \& Shibata 2005; Zanni et al. 2007; Ahmed \& Shibata 2008;
Romanova et al. 2018), in which a sudden switch on of the jet is
definitely not imposed (the ``switching on'' of the outflow being
a result of the simulated accretion+ejection flow). These models
produce a highly superalfv\'enic leading head that evolves consistently
with the computed time-dependence of the ejection.

In the present paper, we revisit the ``slow turn on jet'' problem. We study the
dynamics of the head of a cylindrical jet with a linear ramp of increasing
ejection velocity as a function of ejection time. We combine this ejection velocity
with two forms for the ejection density variability:
\begin{itemize}
  \item a constant ejection density,
  \item a constant mass loss rate (so that the ejection density is proportional
  to the inverse of the ejection velocity).
\end{itemize}
We obtain analytical models of this flow by assuming that the material in the
jet beam is free streaming (before reaching the jet head), and that the motion
of the working surface of the jet head can be described:
\begin{itemize}
\item by using a ram-pressure balance condition,
\item by assuming that it coincides with the motion of the center of
  mass of the material that has entered the working surface.
\end{itemize}
The first of these assumptions (see, e.g., Raga \& Cant\'o 1998) is correct for
a ``massless'' working surface which instantaneously ejects most of the material sideways (into a jet cocoon), while the latter assumption
(see Cant\'o et al. 2000) is appropriate for the ``mass conserving'' case in which
the shocked material mostly stays within the working surface.

We also carry out axisymmetric numerical simulations of the flow, and compare
the results with the ram-pressure balance and center of mass analytic models.
This is the first time that a comparison between numerical simulations
and the two analytic approximations (described above) has been attempted.

The paper is organized as follows. In \S2 we present the center of mass and the
ram-pressure balance solutions for a uniformly accelerated jet evolving in a
homogeneous interstellar medium. In \S3 we present the axisymmetric numerical
simulations, and a comparison of the results with the analytic models is done
in \S4. A discussion of the results is held in \S5, and concluding remarks are
given in \S6.

\section{Analytic model}
\label{sec:a-m}

We consider a jet with an ejection velocity of the form
\begin{equation}
    u_0(\tau) = \left\{\begin{array}{lr}
         0 \hspace{0.1cm}, & \tau<0  \\
         a\tau \hspace{0.1cm} , & \tau \ge 0
    \end{array} \right.
    \label{ejec_vel}
\end{equation}
where $a$ is a constant acceleration and $\tau$ the ejection time.
We assume a cylindrical flow, with a jet cross section $\sigma$ which
is circular and constant. We also assume that the jet moves into an
environment of uniform density $\rho_a$.

We consider two different forms for the ejection density:
\begin{enumerate}
\item a constant mass loss rate per unit area ${\dot m}$,
  such that the ejection density is given by
    \begin{equation}
        \rho_0(\tau) = \frac{\dot{m}}{u_0(\tau)},
        \label{eq:rhotau}
    \end{equation}
    \item a time-independent ejection density,
    \begin{equation}
        \rho_0=\mathrm{const.}
        \label{eq:rhocte}
    \end{equation}
 \end{enumerate}
 
We assume that the flow in the jet beam is free-streaming, satisfying
the condition:
\begin{equation}
    u(x,t)=\frac{x}{t-\tau}=u_0(\tau)\,,
    \label{position}
\end{equation}
where $u(x,t)$ is the velocity as a function of distance $x$ from
the source at an evolutionary time $t$ and $u_0(\tau)$ is the ejection velocity
(at the ejection time $\tau<t$). For an ejection time $\tau<0$, $x=0$,
i.e., no material is ejected. Also, for a cylindrical, free-streaming flow,
the density is given by:
\begin{equation}
    \rho(x,t) = \frac{\rho_0\,u_0}{u_0 - (t - \tau){\dot u}_0},
\label{density}
\end{equation}
where $\rho_0(\tau)$ is the ejection density (see equation \ref{eq:rhotau} or \ref{eq:rhocte})
and ${\dot u}_0=du_0/d\tau$ is the derivative of the ejection velocity
with respect to the ejection time (for a derivation of this equation,
see \citeauthor{RK92}, 1992). The relationship between the evolutionary time $t$ and the ejection time $\tau$ can be obtained from equations (\ref{ejec_vel}) and (\ref{position}). Later, in section 2.1, we explicitly show this relationship.

When the flow starts (at $t=0$), a working surface is formed at $x=0$. This ``jet
head'' then travels away from the outflow source for increasing times. In order
to describe the time evolution of this working surface, we use two analytic
approximations:
\begin{enumerate}
\item \textit{center of mass}- we assume that the position of the working surface
  coincides with the center of mass of the ejected, free-streaming material,
\item \textit{ram-pressure balance}- we assume that the motion of the working
  surface is determined by the jet/environment ram-pressure balance condition.
\end{enumerate}
The first of these approximations is appropriate for a working surface in which
the shocked gas mostly remains within the jet head, and the latter approximation
is valid for a working surface that ejects most of the matter sideways (into
a jet cocoon).

We therefore develop four analytic models, with the two ejection densities
and the two analytic approximations described above. The models all share
the linearly accelerating ejection velocity given by equation (\ref{ejec_vel}).

%
%
 \subsection{Center of mass equation of motion}
 
 We first consider a ``mass conserving'' working surface. For this case,
 we assume that the material going through the bow and jet shocks
 mostly stays within the working surface, with only a small fraction
 of this material being ejected sideways into the jet cocoon.
 The working surface position should then coincide with the center of mass
 of the free-streaming fluid parcels that have piled up within it. This
 center of mass position is given by:
 \begin{equation}
     x_{cm} = \frac{\int xdm}{\int dm},
     \label{center_mass}
 \end{equation}
with the differential element of mass given by:
    \begin{equation}
        dm = \sigma\, \rho_0(\tau)\, u_0(\tau)\, d\tau+\sigma\, \rho_a(x)\, dx.
        \label{dm}
    \end{equation}
    In this equation, the first term on the right-hand-side corresponds
    to the ejected material,
    and the second term to the swept-up, stationary environment entering the
    working surface. In 
    equation (\ref{dm}), $\sigma$ is the jet cross section, $\rho_0(\tau)$ is
the ejection density, $u_0(\tau)$ the ejection velocity and $\rho_a(x)$
the (possibly position-dependent) environmental density.

Using equations~(\ref{position}-\ref{dm}) and considering that the ejection
time $\tau'$ is integrated from 0 to $\tau$, we obtain:
\begin{multline}
     x_{cm} \left[\int_0^\tau \rho_0(\tau')\,u_0(\tau')\,d\tau' + \int_0^{x_{cm}} \rho_a\,dx \right] = \\ \int_0^\tau x_j\,\rho_0(\tau')\,u_0(\tau)\,d\tau' + \int_0^{x_{cm}}x\,\rho_a\,dx,
     \label{center_mass_2}
\end{multline}
where we have set $\rho_a=const.$, and
$x_j$ is the position that the fluid parcels would have if they were
still free-streaming:
\begin{equation}
  x_j = (t-\tau')\,u_0(\tau')\,,
  \label{xjttau}
\end{equation}
see equation (\ref{position}).

Also, from equations (\ref{ejec_vel}) and (\ref{position}) we find:
\begin{equation}
        t=\frac{x_{cm}}{a\tau}+\tau\,.
        \label{timet}
\end{equation}
Equation (\ref{timet}) can be inverted to obtain
\begin{equation}
        \tau = \frac{1}{2}\left(t + \sqrt{t^2 - \frac{4x_{cm}}{a}} \right).
        \label{timetau}
\end{equation}

Combining equations (\ref{center_mass_2}-\ref{timet}), we obtain:
\begin{multline}
  x_{cm} \left[\int_0^\tau \rho_0(\tau')\,a\tau'\,d\tau' + \frac{\rho_a\,x_{cm}}{2}
    \right] = \\
  \int_0^\tau (t-\tau')a^2\,\tau'^2\,\rho_0(\tau')\,d\tau'\,.
     \label{center_mass3}
\end{multline}

In order to carry out the remaining integrals, we have to specify
$\rho_0(\tau')$. We consider two forms for
the ejection density (see equation \ref{eq:rhotau}):

\begin{enumerate}[a)]
\item { \it Constant mass loss rate }
  
equation (\ref{center_mass3}) takes the form:
      \begin{multline}
        x_{cm}\left(\dot{m}\tau + \frac{\rho_a\,x_{cm}}{2} \right) = \dot{m}a\tau^2\left(\frac{t}{2} - \frac{\tau}{3} \right).
        \label{ctemassloss}
    \end{multline}
   Using equations (\ref{timet}) and (\ref{ctemassloss}), we obtain
    \begin{equation}
        x_{cm}^2 + \frac{\dot{m}\tau}{\rho_a}\,x_{cm} - \frac{\dot{m}a\tau^3}{3\rho_a} = 0,
        \label{solution_mlc}
    \end{equation}
this is a quadratic equation with one positive solution, given by
    \begin{equation}
        x_{cm}(\tau) = \frac{\dot{m} \tau}{2\rho_a} \left[-1 + \sqrt{1+\frac{4a\rho_a \tau}{3\dot{m}}} \right].
        \label{mdotcte_solution}
    \end{equation}
   
   \begin{figure*}[ht]
   \centering
   \includegraphics[scale=0.55]{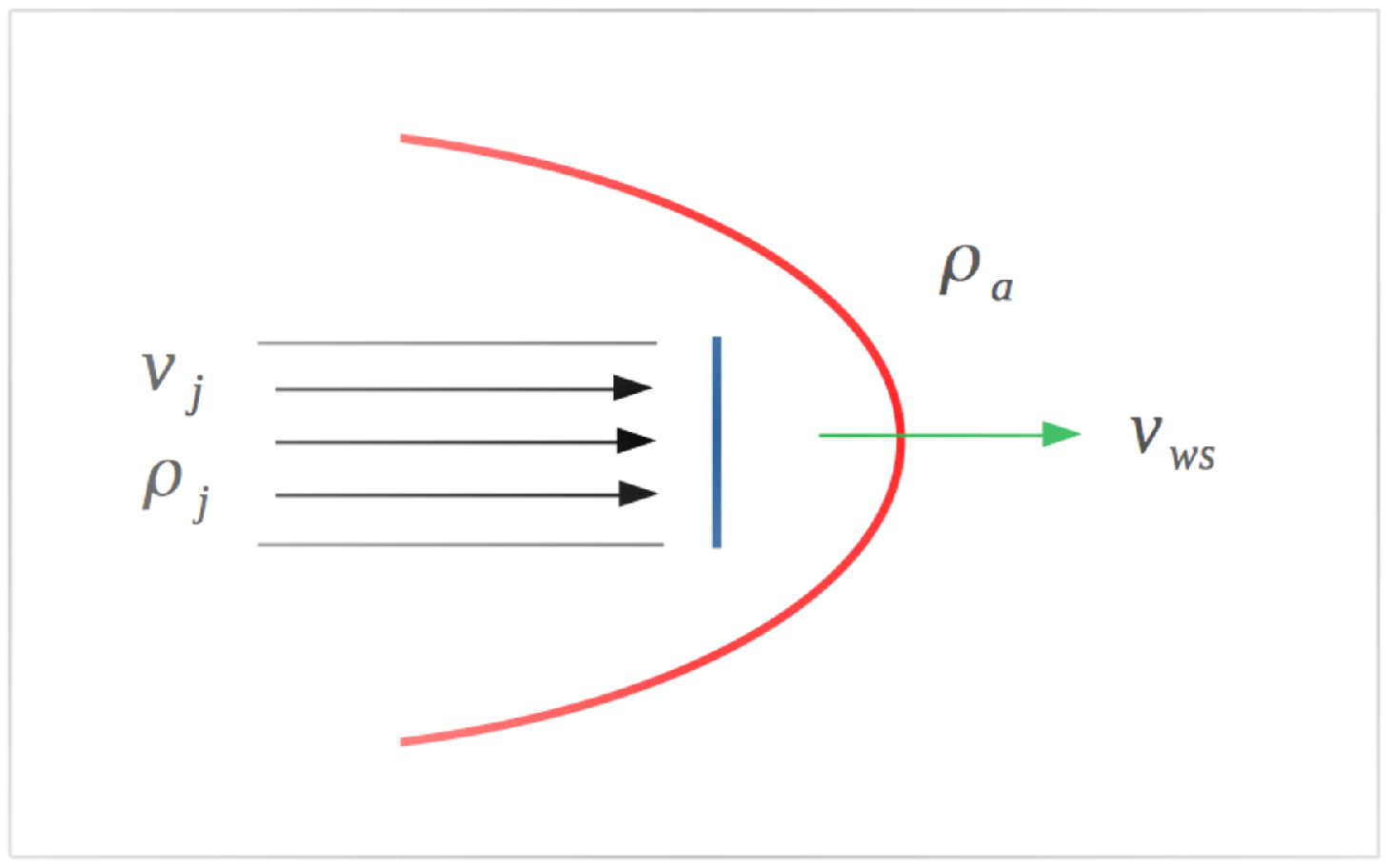}
   \includegraphics[scale=0.55]{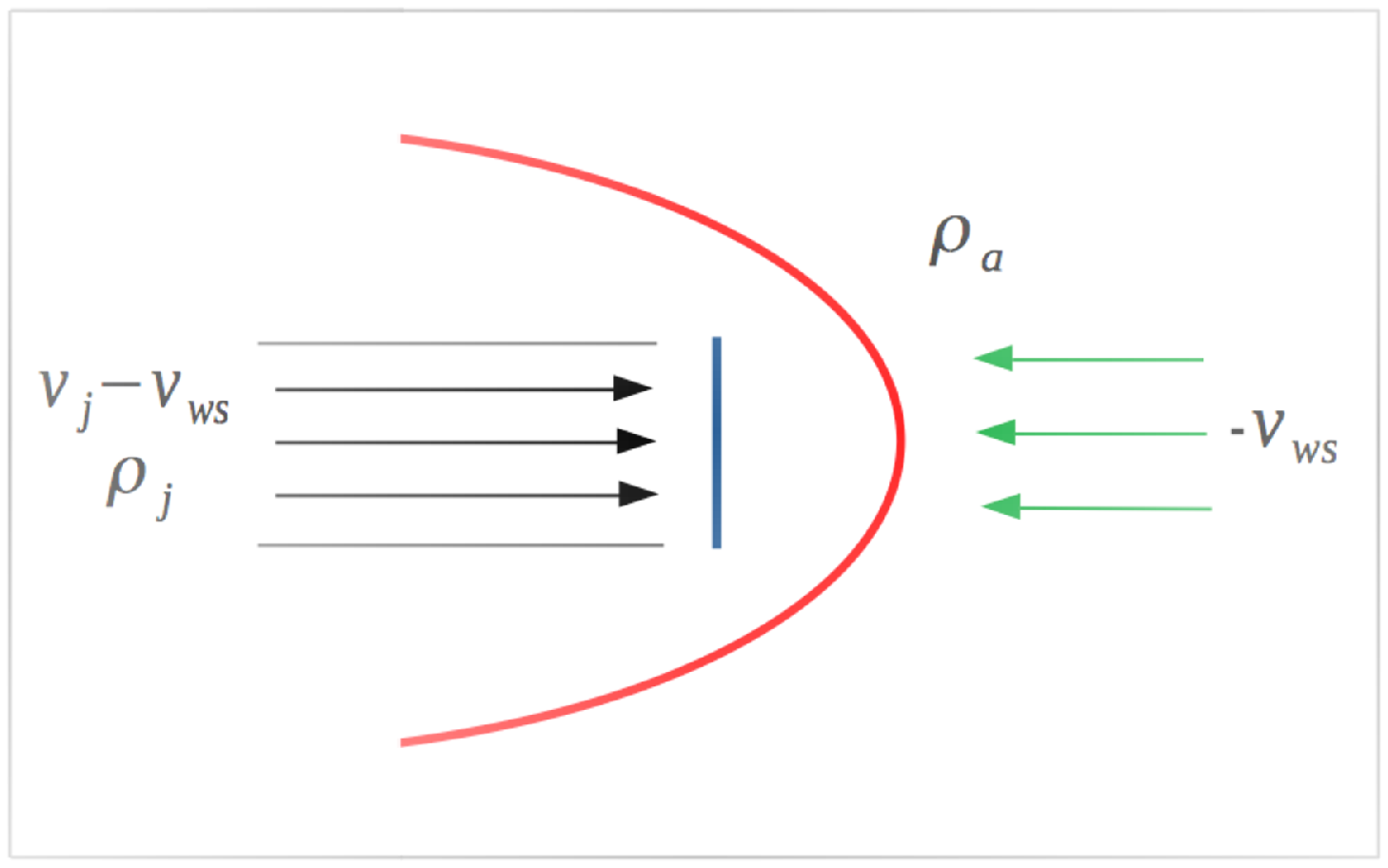}
   \caption{\footnotesize{Jet  head working surface in a reference frame at rest with respect to the outflow source (left) and in a reference frame moving along with the working surface (right). The bow shock is in red and the jet shock is in blue.}} 
   \label{bowshock}%
   \end{figure*}
   
    In the low density limit $\rho_a \rightarrow 0$, equation (\ref{solution_mlc}) takes a particularly simple form that leads to finding the solution
    \begin{equation}
        x_{cm}=\frac{a\tau^2}{3}.
    \end{equation}
This solution can also be obtained by carrying out a first-order
    Taylor series expansion of equation (\ref{mdotcte_solution}).
    
   By substituting equation (\ref{timetau}) into (\ref{ctemassloss}), we can also obtain $x_{cm}$ as an explicit function of $t$, giving
    \begin{equation}
        x_{cm} = \frac{8}{9}x_c \left[\left(1+\frac{3}{4} \frac{t}{t_c} \right)^{3/2} - \left(1+ \frac{9}{8} \frac{t}{t_c} \right) \right],
        \label{sol_xcm_t}
    \end{equation}
   with, $
        t_c \equiv \dot{m}/a\rho_{a}$ and $ x_c \equiv \dot{m}^2/a\rho_a^2$.
        
        In the $t \ll t_c$ limit, a second order Taylor series
          expansion of equation (\ref{sol_xcm_t}) gives:
    \begin{equation}
        x_{cm} \approx \frac{3}{16} at^2.
    \end{equation}

    \vspace{0.5cm}
  \item { \it Constant ejection density }
    
With a constant $\rho_0$, equation (\ref{center_mass3}) takes the form:
    \begin{multline}
        x_{cm}\left(\frac{\rho_0\,a\tau^2}{2} + \frac{\rho_a\,x_{cm}}{2} \right) = \\ \rho_0\,a^2\tau^3 \left(\frac{t}{3} - \frac{\tau}{4} \right)\,,
        \label{eqmotion_rhoc}
    \end{multline}
where $t$ and $\tau$ are related through equation (\ref{timet}).
    
Substituting $t$ as a function of $\tau$ we obtain:
    \begin{equation}
      x_{cm}^2 + \frac{\rho_0\,a\tau^2}{3\rho_a}x_{cm} -\frac{\rho_0\,a^2\tau^4}{6\rho_a}=0,
      \label{xxxcm}
    \end{equation}
    which has a positive solution
    \begin{equation}
        x_{cm}(\tau) = \frac{\rho_0\,a\tau^2}{6\rho_a}\left[-1 + \sqrt{1+\frac{6\rho_a}{\rho_0}} \right].
        \label{solution_rhoc}
    \end{equation}
    In the $\rho_a \rightarrow 0$ limit, equation (\ref{xxxcm}) (or,
    alternatively, equation \ref{solution_rhoc}) takes the form
    \begin{equation}
        x_{cm}=\frac{a\tau^2}{2},
    \end{equation}
    with a quadratic dependency on the ejection time.
    
    We can also find a solution to (\ref{eqmotion_rhoc}) as a function of
    evolutionary time $t$ (using equation \ref{timetau}), giving:
    \begin{equation}
        x_{cm} = \frac{a}{9}\beta_0 \left[\frac{\beta_0 (\beta_0^2 -18) + (\beta_0^2 + 6)^{3/2}}{(\beta_0^2 - 2)^2} \right] t^2,
        \label{solution_timet}
    \end{equation}
    where,
    \begin{equation}
        \beta_0 = \sqrt{\frac{\rho_0}{\rho_a}},
        \label{beta0}
    \end{equation}
    and therefore the position $x_{cm}$ has a quadratic dependency on
    the evolutionary time $t$.
    
    Equation (\ref{solution_timet}) takes the form of a constant acceleration condition, with the acceleration given by
    \begin{equation}
        g = \frac{2a}{9}\beta_0 \left[\frac{\beta_0 (\beta_0^2 -18) + (\beta_0^2 + 6)^{3/2}}{(\beta_0^2 - 2)^2} \right],
    \end{equation}
    which in the $\rho_a \rightarrow 0$ low density limit becomes,
    \begin{equation}
        g \approx \left(\frac{2}{3} \right)^{1/2} a\, \beta_0.
    \end{equation}

\end{enumerate}

\subsection{Ram-pressure balance equation of motion}

Figure \ref{bowshock} shows a schematic diagram of a jet head propagating
through a stationary environment. This leading working surface has
a structure consisting of two shocks: an upstream shock known as the
jet shock (or Mach disk), which is slowing down the jet material, and a downstream
bow shock which is accelerating the surrounding material. The two-shock
working surface structure travels away from the outflow source at a velocity
$v_{ws}$.

The right panel of Figure \ref{bowshock} shows the situation seen in
a reference frame at motion with the jet head. For a hypersonic flow, the
two working surface shocks are strong, so that the post-shock gas pressures
are given by:
\begin{equation}
    P_{bs}=\frac{2}{\gamma+1}\rho_a v_{ws}^2,
\end{equation}
for the bow shock, and
\begin{equation}
    P_{js}=\frac{2}{\gamma+1}\rho_j\,(v_j - v_{ws})^2,
\end{equation}
for the jet shock (assuming that the jet and the environment have
the same specific heat ratio $\gamma$). In this equation,
$v_j$ and $\rho_j$ are the velocity (with respect to the outflow
  source) and the density of the material that is presently entering
  the jet shock.

If the working surface is moving with a constant velocity then the condition
\begin{equation}
     P_{bs} = P_{js}\; \Rightarrow\;  \rho_a v_{ws}^2 = \rho_j\,(v_j - v_{ws})^2,
     \label{rampressure_balance}
\end{equation}
is satisfied. This is called the 
\textit{ram-pressure balance} condition. This condition is also valid for a
working surface moving with a variable velocity as long as the inertia of
the material between the bow shock and the jet shock is negligible, which is
the case if most of the material is ejected sideways from the working surface
into a jet cocoon.

From equation (\ref{rampressure_balance}) we find that 
\begin{equation}
  v_{ws} = \frac{\beta v_j}{1+\beta}\hspace{0.2cm},\;\hspace{0.5cm}
  {\rm with} \hspace{0.5cm}\beta \equiv \sqrt{\frac{\rho_j}{\rho_a}}.
\end{equation}

Clearly, the shock velocity associated with the bow shock is $v_{bs}=v_{ws}$ and
the shock velocity of the jet shock is:
\begin{equation}
    v_{js}=\frac{v_j}{1+\beta}.
\end{equation}

We now consider the equation of motion $v_{ws}=dx_{ws}/dt$ for the working
surface. Setting $v_j=u_0(\tau)$ (i.e., the velocity of the material entering
the working surface is equal to the ejection velocity at the corresponding
ejection time $\tau$), we then have:
\begin{equation}
    \left(1+\sqrt{\frac{\rho_a}{\rho_j}} \right)\frac{dx_{ws}}{dt} = u_0(\tau).
    \label{motion_rampress}
\end{equation}
The ejection time $\tau$ of the material entering the working surface is obtained
from the free-streaming relation
\begin{equation*}
    x_{ws}=(t-\tau)u_0(\tau),
\end{equation*}
where $t$ is the evolutionary time (see equation  \ref{position}).
Solving for $t$ and differentiating
with respect to $\tau$ we obtain
\begin{equation}
    \frac{dt}{d\tau} = 1 + \frac{1}{u_0}\frac{dx_{ws}}{d\tau} - \frac{x_{ws}}{u_0^2}\frac{du_0}{d\tau},
    \label{diff_times}
\end{equation}
Combining equations (\ref{motion_rampress}) and (\ref{diff_times}), we obtain:
\begin{equation}
    \frac{dx_{ws}}{d\tau}\sqrt{\frac{\rho_a}{\rho_j}} = u_0 -x_{ws}\frac{d\ln{u_0}}{d\tau}.
    \label{diffeqws}
\end{equation}
The density of the jet material entering the working surface is
$\rho_j=\rho(x_{ws},t)$ (see equation \ref{density}), which for our chosen
ejection velocity variability (see equation \ref{ejec_vel}) becomes:
\begin{equation}
    \rho_j= \frac{\rho_0}{1-\frac{x_{ws}}{a\tau^2}}.
    \label{dens_tau_x}
\end{equation}
Finally, from equations (\ref{diffeqws}) and (\ref{dens_tau_x}) we obtain:
\begin{equation}
  \frac{dx_{ws}}{d\tau}= \sqrt{\frac{\rho_0}{\rho_a}}\,a\tau\,
  \sqrt{1-\frac{x_{ws}}{a\tau^2}}.
    \label{diffeqwsrhoc}
\end{equation}

In order to proceed it is now necessary to specify the form of $\rho_0(\tau)$.
We therefore consider the two cases of constant mass loss rate and constant
ejection density.

\begin{enumerate}[a)]
\item { \it Constant mass loss rate }
  
   for $\dot{m}=\mathrm{const.}$ equation (\ref{diffeqwsrhoc}) takes the form:
    \begin{equation}
        \frac{dx_{ws}}{d\tau} = \sqrt{\frac{\dot{m}}{\rho_a} \left(a\tau - \frac{x_{ws}}{\tau}\right)}.
        \label{ctemasslosseqqdiff}
    \end{equation}
    It is convenient to use the dimensionless variables:
    \begin{equation}
        \eta = \frac{a\rho_a^2}{\dot{m}^2}x_{ws} \;,\;\;\;y=\frac{a\rho_a}{\dot{m}}\tau.
        \label{dimensionless_var}
    \end{equation}
    In terms of these variables, equation (\ref{ctemasslosseqqdiff}) is
    \begin{equation}
        \frac{d\eta}{dy} = \sqrt{y - \frac{\eta}{y}}.
        \label{dimensionless_eqmov}
    \end{equation}
    In Appendix A, it is shown that an approximate analytic solution
    of equation (\ref{dimensionless_eqmov}) can be constructed as a non-linear
    average of a ``near'' and a ``far field'' analytic solution
    (see equations \ref{near_far} and \ref{average}),
    which in terms of the respective dimensional variables is
   \begin{equation}
     x_{ws} = \frac{\dot{m}^2}{a \rho_a^2} \left[\left(\frac{a\rho_a}{\dot{m}}\tau \right)^{-\frac{5}{2}} +\left(\frac{3}{2}\right)^{\frac{5}{4}} \left(\frac{a\rho_a}{\dot{m}}\tau\right)^{-\frac{15}{8}}\right]^{-\frac{4}{5}}.
     \label{xwsolm}
   \end{equation}

    \vspace{0.5cm}
  \item { \it Constant ejection density }
    
    Setting $\rho_0=\mathrm{const.}$ in (\ref{diffeqwsrhoc}) and taking the
      square of the equation, we obtain:
    \begin{equation}
        \frac{\rho_a}{a\rho_0} \left(\frac{dx_{ws}}{d\tau} \right)^2 + x_{ws} - a\tau^2 = 0.
    \end{equation}
    Proposing a power law solution, we find that
    \begin{equation}
      x_{ws} = \frac{a\rho_0}{8\rho_a} \left[-1 + \sqrt{1+ \frac{16 \rho_a}{\rho_0}} \right] \tau^2.
      \label{xxxws}
    \end{equation}
    We therefore find a quadratic dependency on the ejection time.
    
    Using equations (\ref{timetau}) and (\ref{xxxws}) we then obtain:
    \begin{equation}
        x_{ws} = \frac{8a\beta_0^2 \left( -1+ \sqrt{1+\frac{16}{\beta_0^2}}\right)}{\left[8 + \beta_0^2 \left(-1 + \sqrt{1+\frac{16}{\beta_0^2}}\right) \right]^2}\,t^2,
        \label{solution_ws_time}
    \end{equation}
    where $\beta_0$ is given by equation(\ref{beta0}).
    This implies a quadratic dependence on the evolutionary time. 
    
    In the $\beta_0 \ll 1$ limit, equation (\ref{solution_ws_time}) becomes
    \begin{equation}
        x_{ws} \approx \frac{a}{2}\,\beta_0\,t^2,
    \end{equation}
    and the $\beta_0 \gg 1$ limit leads to
    \begin{equation}
        x_{ws} \approx \frac{a}{4}\,t^2.
    \end{equation}
    
\end{enumerate}

\section{The numerical simulations}
\label{sec:n-s}
In order to check our analytical solutions we computed a set of 2D axisymmetric simulations for both the constant mass loss rate and constant ejection density cases.
For the simulations we use a code that solves the ideal gas-dynamic (Euler) equations in a fixed two dimensional grid. The code uses a second order Godunov type method with the HLLC \citep{Toro94} approximate Riemann solver, including a linear reconstruction of the primitive variables with the {\it minmod} slope limiter to avoid spurious oscillations. In order to use cylindrical coordinates, the appropriate geometrical source ($\propto 1/r$) terms are included after each timestep in an operator splitting fashion.
Additionally, we incorporated the
  cooling function dependent on density, metallicity and temperature in the energy equation (also as source term) as proposed by \citet{2014MNRAS.440.3100W}, which we compute assuming
  a solar metallicity.

  In order to stabilize the method an artificial diffusion with a (dimensionless)
  value of $0.01$ was added to all of the equations, and a Courant number
  of $0.2$ was used. The code is written in fortran90
  and is paralelized with the {\it  Message Passing Interface.}
We used a computational grid with a size of 0.05 and 0.2 pc along the $r$ and $x$ directions, respectively. The spatial resolution was $\sim 6.89\,\mathrm{au}$, corresponding to $600\times 6000$ cells along the radial and axial directions.

\subsection{Initial and boundary conditions}

We model jets propagating into a uniform, quiescent environment, and consider
the values $n_{a}=100\,$ cm$^{-3}$ and $n_{a}=5000\,$ cm$^{-3}$ for
the environmental density. For convenience, from now on we will use number density instead of mass density. We consider a mean molecular weight of $\mu =1.3$. We also consider the cases of jets with a constant
mass loss rate and a constant ejection density.

In all simulations, we consider the ejection velocity variability
given by equation (\ref{ejec_vel}) with $a= 100\,\mathrm{km\,s^{-1}}$
per millenium$\;=3.17\times10^{-4}\,\mathrm{cm\,s^{-2}}$.
An initial jet radius $r_j=300\,\mathrm{AU}$ (the cross section
being $\sigma=\pi r_j^2$) and a temperature
$T=100\,\mathrm{K}$ (for both the jet and the environment) was imposed
in all models.

We therefore compute four models, two with constant ejection
density (which we label $n_{100}$ and $n_{5000}$, with the subscript
giving the ambient density in cm$^{-3}$) and two with constant
mass loss rate (which we label ${\dot m}_{100}$ and ${\dot m}_{5000}$).
For the models $\dot{m}_{100}$ and $\dot{m}_{5000}$ we choose a total
mass loss rate $\dot{M}=2.24\times10^{-8}\,\mathrm{M_{\odot}yr^{-1}}$,
which corresponds to a mass loss rate at the jet source per unit
area $\dot{m} = 3.36\times 10^{-13}\,\mathrm{g\,cm^{-2}\,s^{-1}}$.
For the models $n_{100}$ and $n_{5000}$, we choose a density
$n_j=1.66\times10^4\,$ cm$^{-3}$ (for a gas with 90\% H and 10\% He).
These model parameters are summarized in Table~\ref{tab:1}.

In the constant mass flux cases, we find a maximum jet density of $n_j=1.56\times 10^7\,\mathrm{cm^{-3}}$, and a minimum jet density of $n_j=6.24\times 10^3\,\mathrm{cm^{-3}}$, corresponding to an ejection time $\tau=1\,\mathrm{yr}$, and $\tau=2500\,\mathrm{yr}$, respectively.

\begin{table}[htbp]
    \centering
    \caption{\footnotesize{Physical conditions for the numerical simulations.}}
    \label{tab:1}
    \begin{tabular}{c|c|c|c}
    \toprule
    & $\dot{M_j}$ & $n_j$ & $n_a$ 
    \\
    &\scriptsize$\mathrm{(M_{\odot}\,yr^{-1})}$ &\scriptsize$\mathrm{(cm^{-3})}$& \scriptsize$\mathrm{(cm^{-3})}$ \\
    \hline
    \hline
    $n_{100}$ & \multirow{2}{*}{$1.30\times 10^{-7^{\,*}}$}  & \multirow{2}{*}{$1.66\times 10^{4}$}  & $_{100}$ \\
    $n_{5000} $ & & & $_{5000}$ \\
    \cline{1-4}
    $\dot{m}_{100}$ & \multirow{2}{*}{ $3.35\times 10^{-7}$}  & \multirow{2}{*}{$1.55\times 10^{4^{\,*}}$} & $_{100}$  \\
    $\dot{m}_{5000}$  &  & & $_{5000}$ \\
    \bottomrule
    \end{tabular}
    \\
    \vspace{0.25cm}
    \footnotesize{$^*$ These values correspond to an ejection
      time $\tau=1000\,\mathrm{yr}$.}
\end{table}

We should note that
  the parameters we have chosen are inspired in the characteristics
  of ``classical'' HH jets such as HH~34 and HH~111, which have:
  \begin{itemize}
    \item  radii $\sim 100$~AU (Reipurth et al. 2002) and $\sim 300$~AU
      (Reipurth et al. 1997), for HH~34 and HH~111, respectively.
      We chose the larger of these
  two radii in order to have a better resolution of the jet in
  our numerical simulations. These jet radii are measured
  in the ``jet knot chain'' at distances of $\sim 10''$ (corresponding
  to $\sim 6\times 10^{16}$~cm) from the outflow sources, and
  not in the considerably broader HH~34S and HH~111V ``heads'',
\item full spatial velocities (i.e., combining radial velocities
  and proper motions) between 200 and 300~km~s$^{-1}$
  (Hartigan et al. 2001; Reipurth et al. 2002) and lengths (out to
  HH~34S and HH~111V) of $\approx 8\times 10^{17}$~cm. We have then made
  a choice of $a= 100\,\mathrm{km\,s^{-1}}$ per millenium (see above),
  which after an evolutionary time of $\sim 2500$~yr produces a
  jet head with a position and velocity similar to the ones of
  HH~34S and HH~111V (see the follwing section). The simulations
  are stopped at this time (i.e., at 2500~yr),
\item mass loss rates in the
  $10^{-8}\to 10^{-7}$~M$_\odot$~yr$^{-1}$ range (depending on the emission
  lines used for deriving this value and on position along the jets,
  see Podio et al, 2006). We have therefore chosen densities
  that produce mass loss rates of this order of magnitude,
\item even though the ambient densities in HH 34 and HH 111 are of $\sim 10$~cm$^{-3}$ (Raga et al. 1991), we have chosen higher environmental densities
  (of 100 and 5000~cm$^{-3}$, see above) in order to have a substantial
  braking effect, and to obtain larger differences between
  the centre of mass and ram-pressure balance solutions.
\end{itemize}
  
\section{Results}

\begin{figure*}[htbp]
    \centering
    \includegraphics[scale=1]{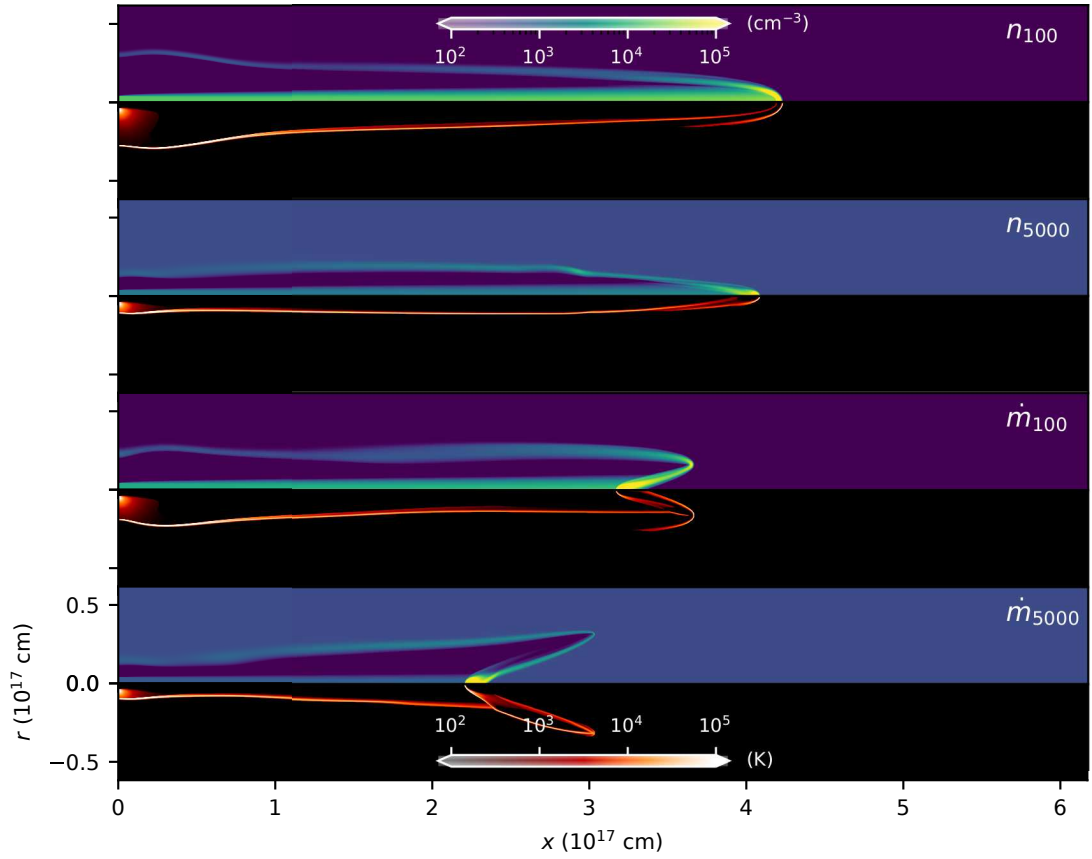}
    \caption{\footnotesize{Numerical density and temperature stratifications obtained from our four models (see the text and Table~1), at an evolutionary time of 2500 years. For each model, we show the number density (top half) and temperature (bottom half) in logarithmic color scales.} The axial and radial axes are given in units of $10^{17}$ cm. In the constant mass loss rate models the jet density, and pressure, drops quite considerably leading to the formation of internal ``crossing shocks''.}
    \label{Figure_2}
\end{figure*}

We computed four numerical simulations,  two with a constant ejection density (models $n_{100}$ and $n_{5000}$) and two with a constant mass loss rate (models $\dot{m}_{100}$ and $\dot{m}_{5000}$). The top (purple background) and bottom (black background) subpanels of each subfigure presented in Figure \ref{Figure_2} display the numerical density and temperature stratifications at an evolutionary time of 2500 yr for these four models. 



Generally, the highest densities and temperatures are found in the
on-axis regions of the jet heads. The jet heads of the models with
constant ejection density (models $n_{100}$ and $n_{5000}$)
have traveled to larger distances from the outflow source than
the ones of constant mass loss rate ($\dot{m}_{100}$ and $\dot{m}_{5000}$).
This qualitative difference is due to the fact that in the constant
mass loss rate models the injected momentum rate ${\dot M}_jv_j$
scales linearly with $v_j$ (and hence also increases linearly
with time, see equation \ref{ejec_vel}), while this scaling is
quadratic (with ejection velocity and with time) in the constant
ejection density models. This means that the constant ejection density jets have a higher momentum content.

As a result of the lower jet beam density and pressure, the constant
mass loss rate models develop an incident/reflected crossing shock
structure, which (in the incident shock region)
leads to the production of faster off-axis motions
in the jet head (see the two bottom panels of Figure~\ref{Figure_2}).
This kind of structure is always found in jet simulations with appropriate parameters, i.e. simulations where the jet is under-dense and under-pressured with respect to the environment during its evolution (see, e.g., Raga 1988; Downes \& Ray 1999).


\begin{figure*}[htbp]
    \centering
    \includegraphics[scale=0.55]{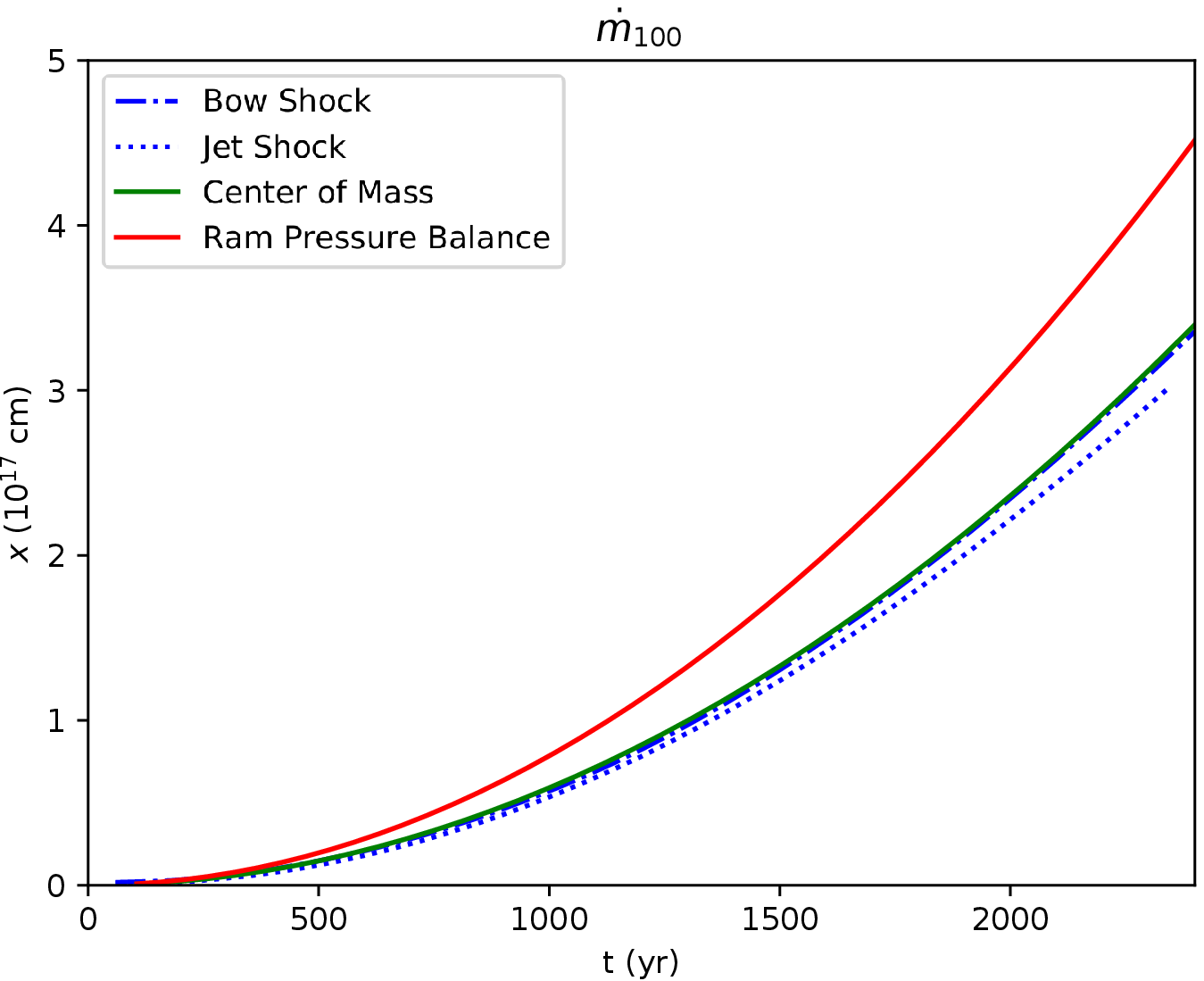}
    \includegraphics[scale=0.55]{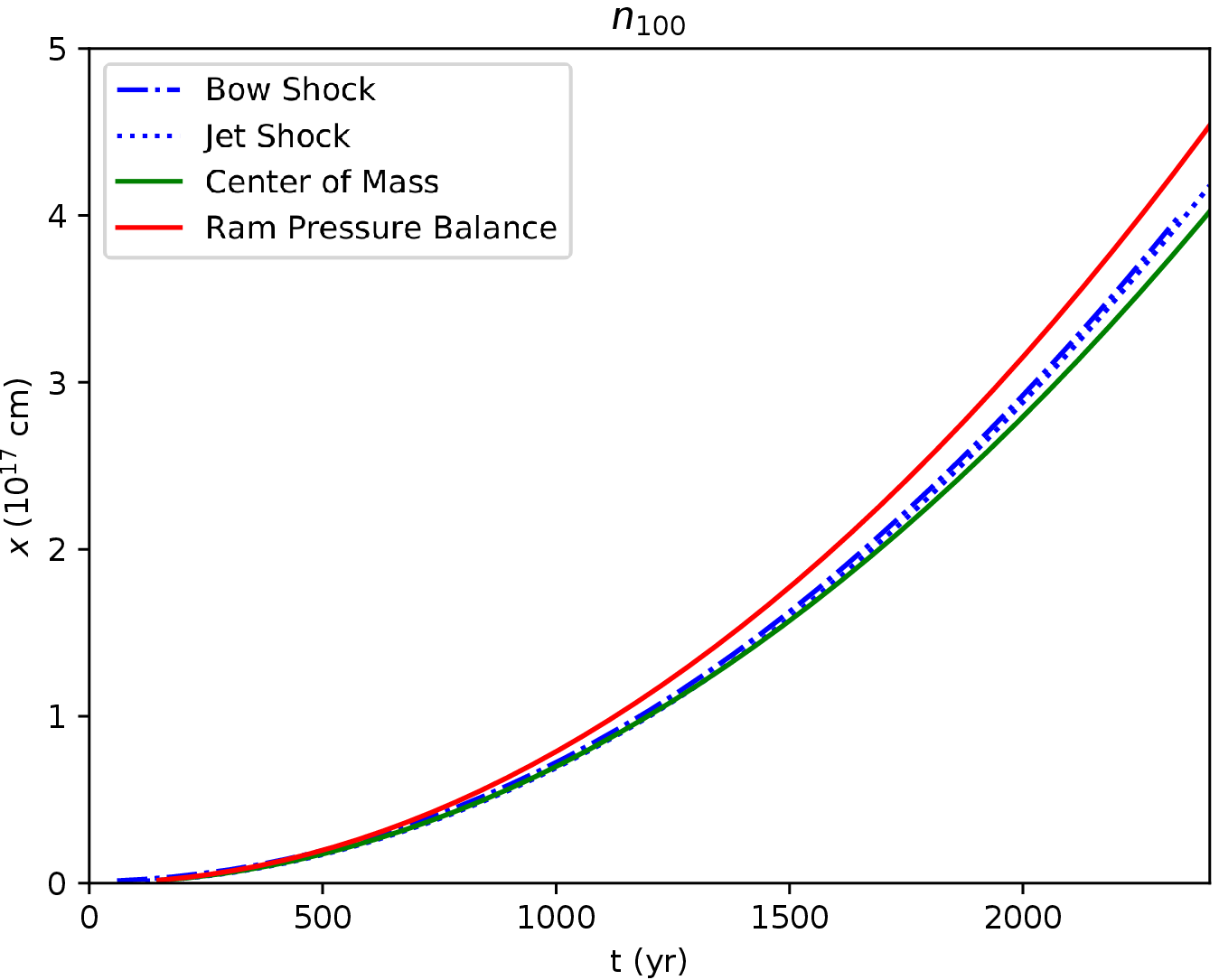}
    \includegraphics[scale=0.55]{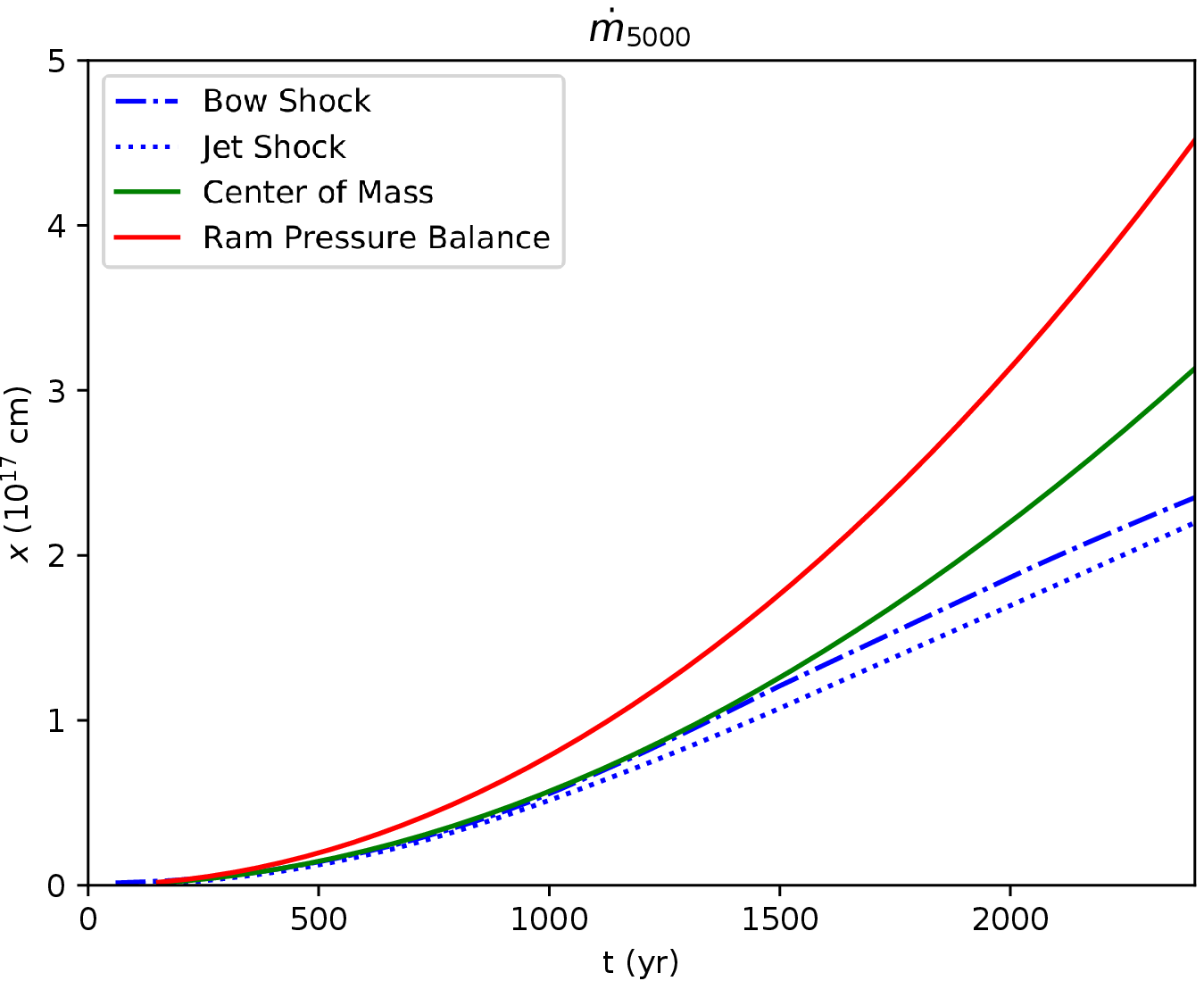}
    \includegraphics[scale=0.55]{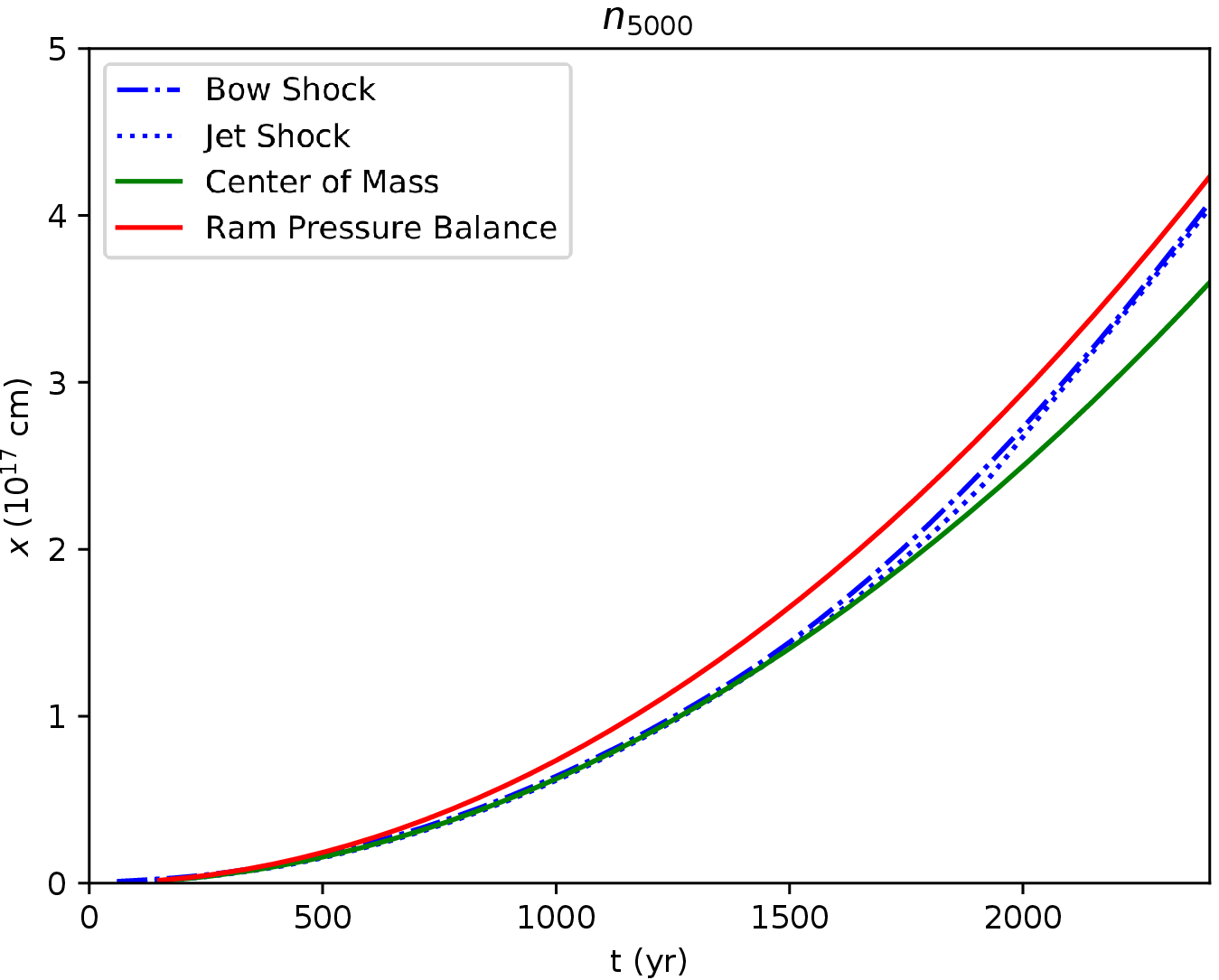}
    \caption{\footnotesize{Position of the jet head as a function of time for the constant mass loss rate models (left panels) and the constant ejection density models (right panels). The red solid line shows the ram-pressure balance solution and the green dashed line shows the center of mass solution. The dash-dotted line shows the bow shock and the dotted line the jet shock positions obtained from the numerical simulations.}}
    \label{posvstime}
\end{figure*}


The upper left panel of Figure~\ref{posvstime} shows the position of the bow shock (dash-dotted blue line) and jet shock (dotted blue line) as a function of time for the ${\dot m}_{100}$ constant mass loss rate model. These positions were obtained by searching for the jumps along the symmetry axis of the pressure stratifications for both the bow shock and jet shock. We compare these shock positions with the analytical solution of the center of mass (dashed green line) and ram-pressure balance (solid red line) analytic models. We find that the position of both the bow shock lies
close to the prediction from the center of mass analytic
model, and the jet shock position lies somewhat below.




The upper right panel of Figure \ref{posvstime} shows the time-dependent position of the jet head for the $n_{100}$ constant ejection density model. For times $<1500$~yr, the jet and bow shock positions (obtained from the numerical simulation) quite closely follow the center of mass analytic solution, At larger times, the jet and bow shock positions lie above the center of mass solution, and begin to approach the ram-pressure balance solution. At all times, the positions of the jet and bow shock lie in between the working surface positions predicted by the two analytic models.


The lower left panel of Figure~\ref{posvstime} shows the evolution of the jet head for the ${\dot m}_{5000}$ constant mass loss rate model. For this model, the positions of the jet and bow shock initially follow the center of mass analytic solution, but for times $t>1500$~yr begin to show large deviations, moving slower than the predictions of the two analytic models. These strong deviations are not surprising given the rather extreme departures from a constant cross section, cylindrical jet beam (assumed in the analytic models) found in the numerical simulation (see the bottom panel of Figure 2).

The strong departures from the simple, cylindrical structure assumed in the analytic models occurs because at increasing times the jet density (and pressure) drop quite considerably in this model, leading to the formation of internal ``crossing shocks''. These shocks then lead to the formation of complex off-axis structures in the jet head (see the two bottom panels of Figure \ref{Figure_2}).



The lower right panel of Figure \ref{posvstime} shows the dynamics of the jet head for the $n_{5000}$ constant ejection density model. This model shows jet and bow shock positions that initially follow the center of mass analytic model, and at times $t>1500$~yr tend to approximate the ram-pressure balance model.



\begin{figure*}[htbp]
    \centering
    \includegraphics[scale=0.55]{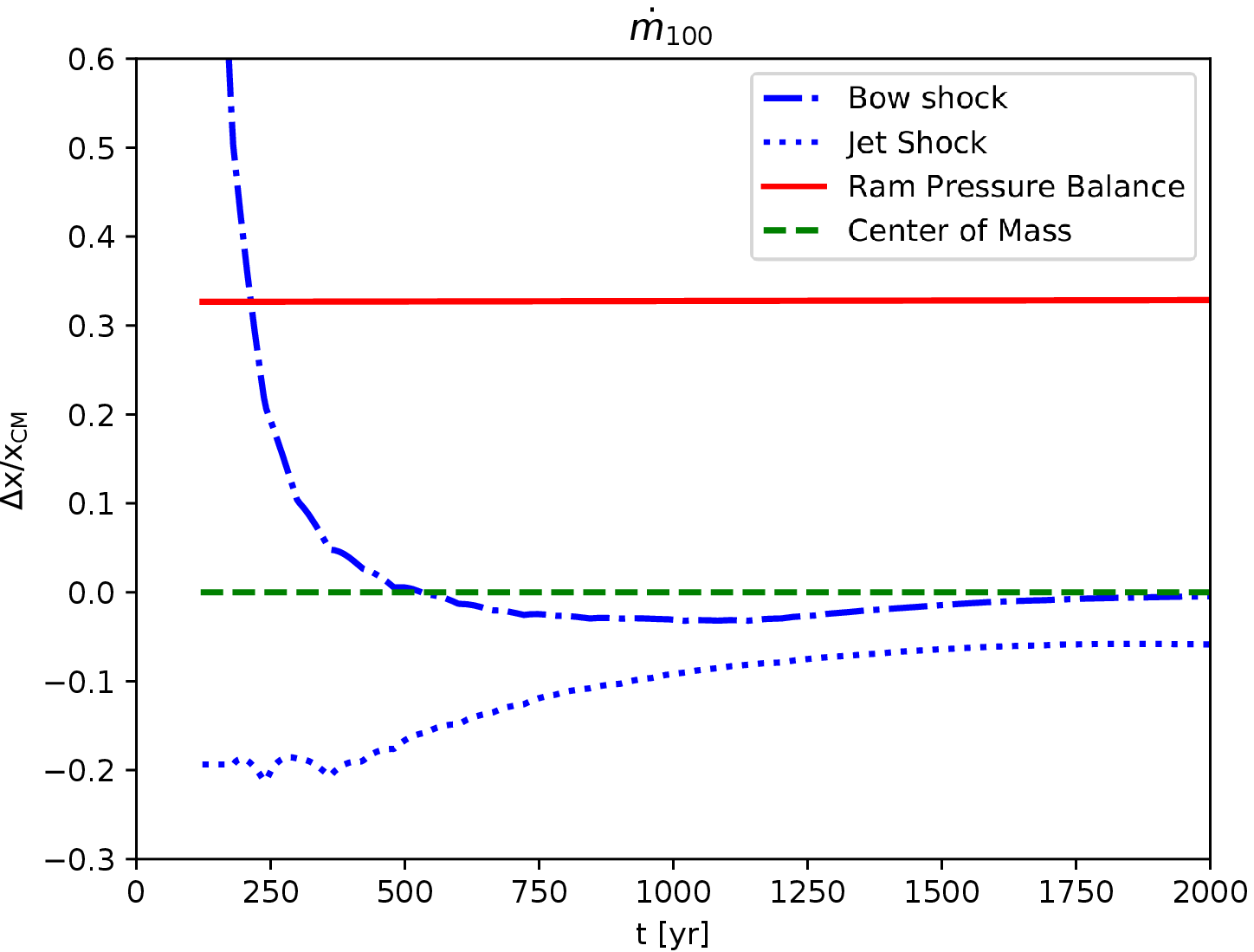}
    \includegraphics[scale=0.55]{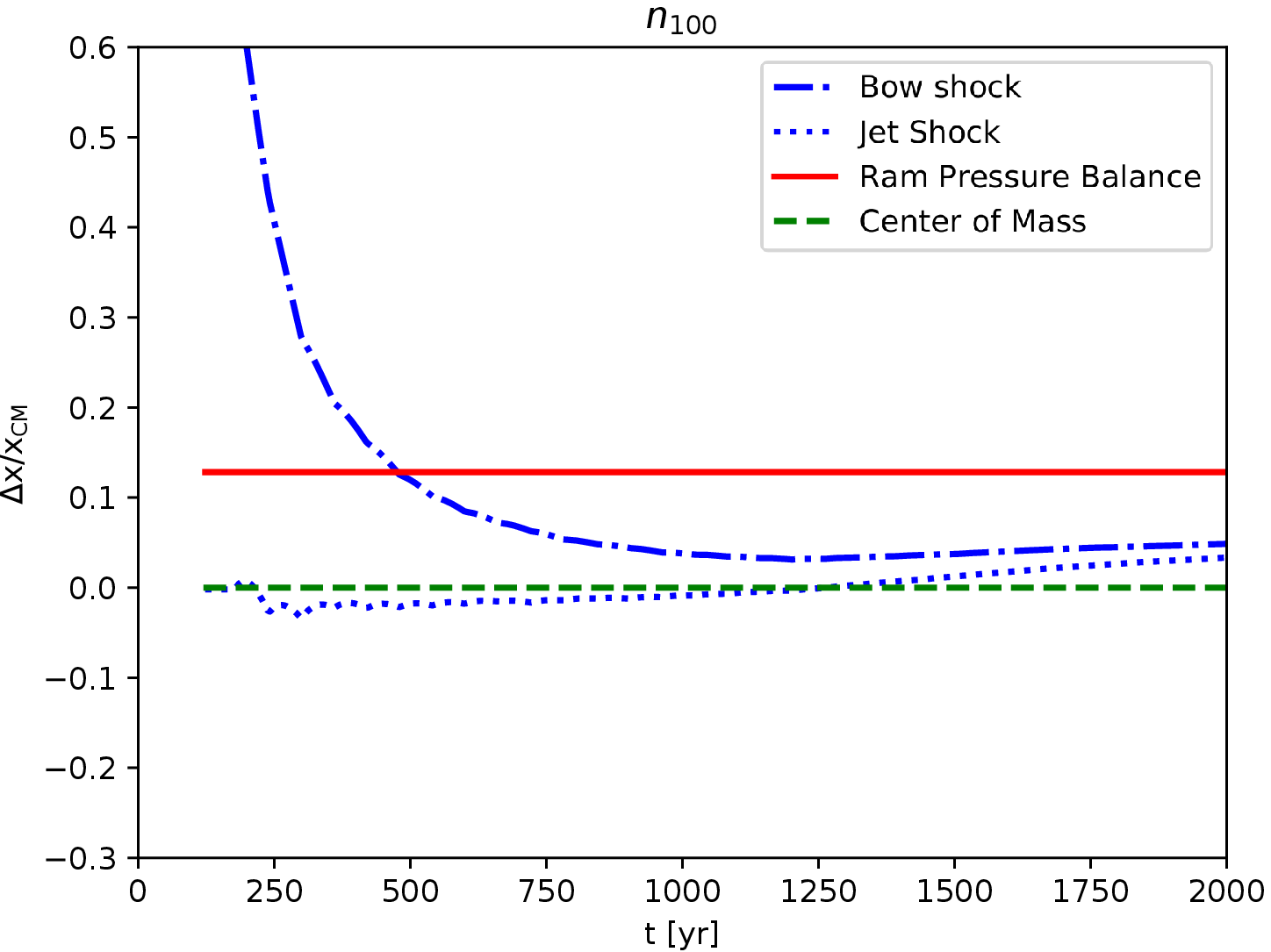}
    \includegraphics[scale=0.55]{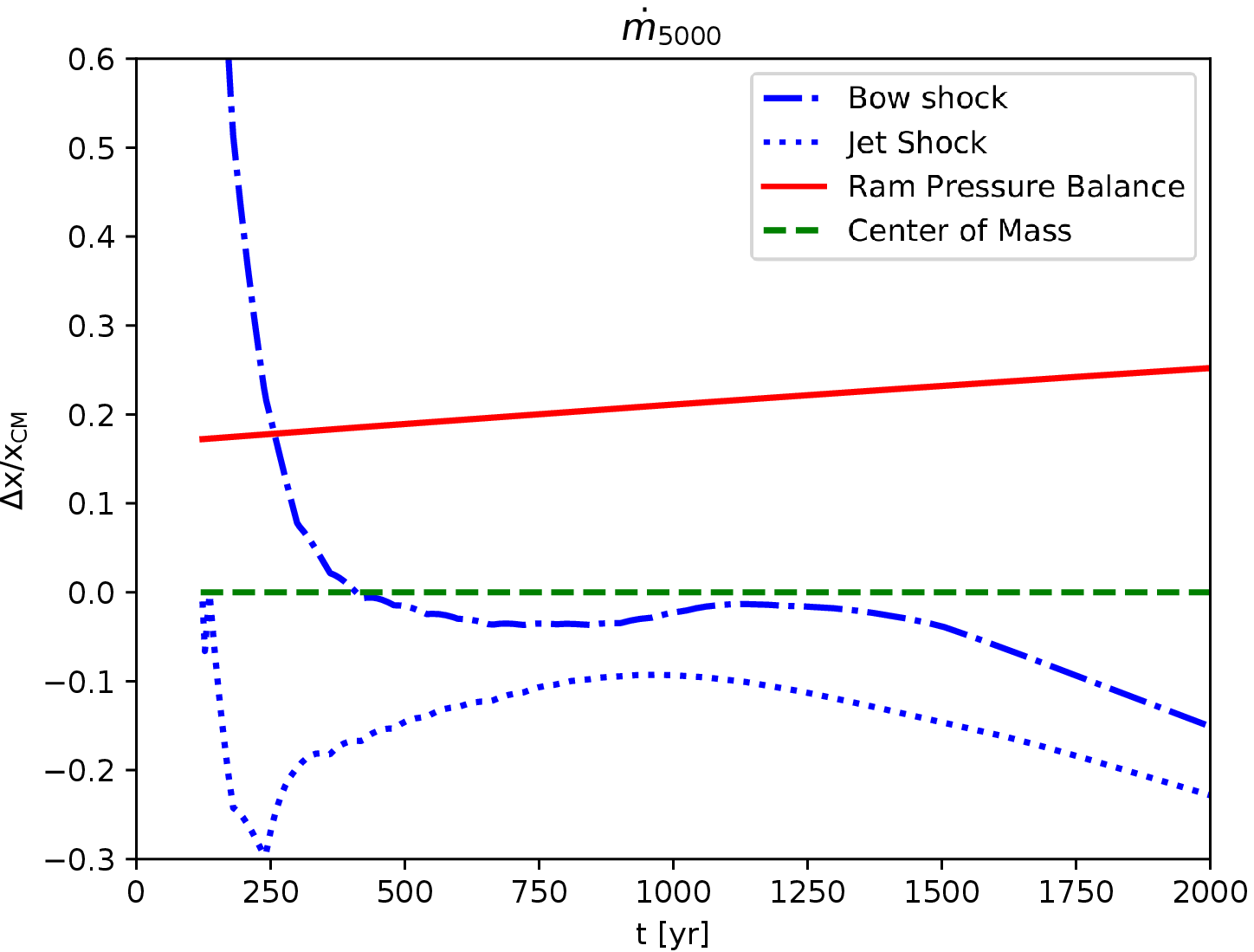} 
    \includegraphics[scale=0.55]{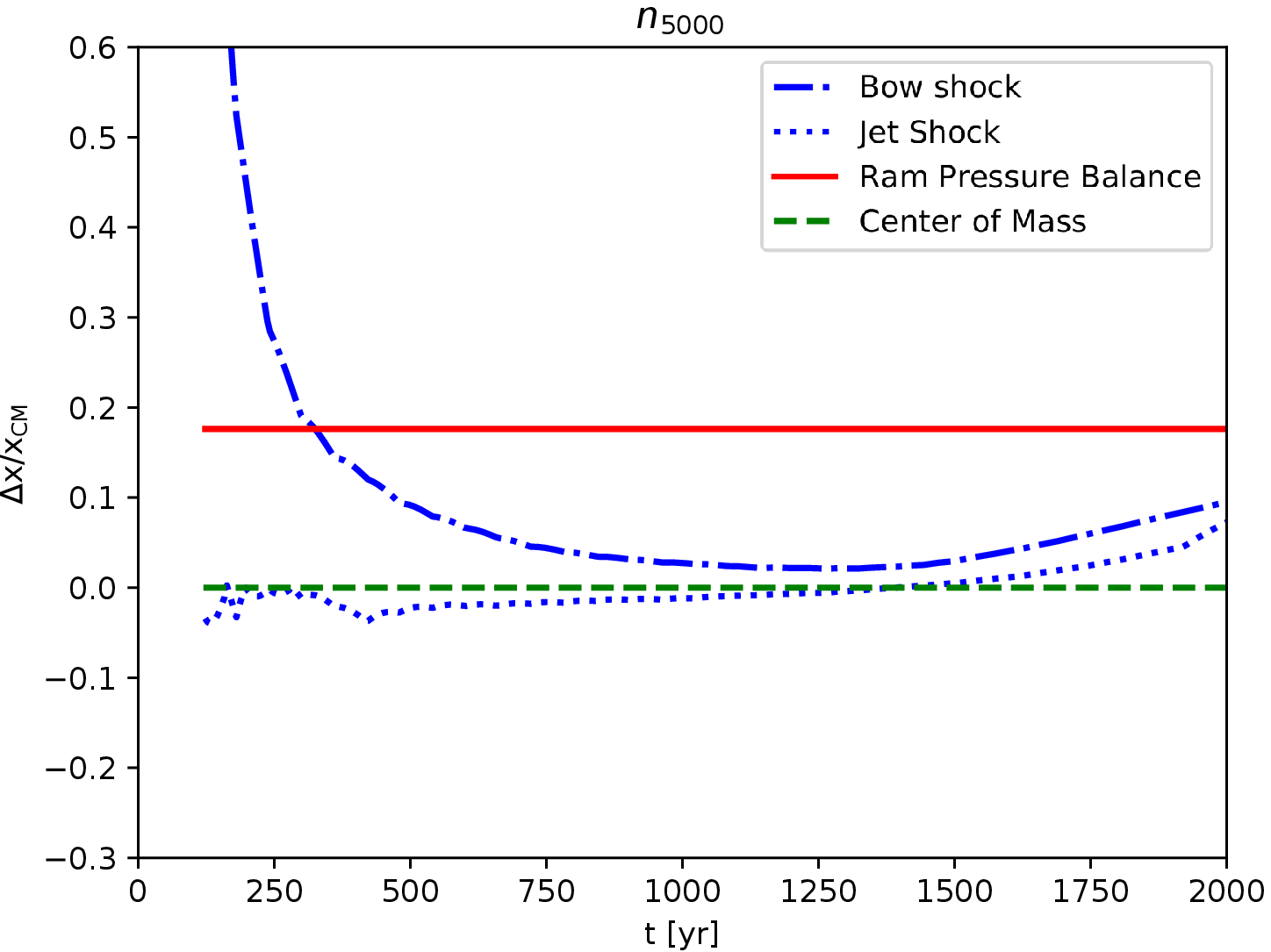}
    \caption{The relative differences of the ram-pressure balance
      solution (solid red line), the numerical bow shock (dash-dotted
      blue line) and jet shock positions (dotted blue line) with
      respect to the center of mass working surface solution (the dashed,
      green line showing the $\Delta x/x_{cm}=0$ line) as a function
      of time $t$. The results obtained for our four models (see Table 1)
      are shown (constant mass loss rate models on the left, and constant
      ejection density models on the right).}
 \label{fig:difverl}
\end{figure*}

Finally, we calculate the relative position differences $\Delta x_a/x_{cm}=
  x/x_{cm}-1$, with $x_{cm}$ being the position of the ``center of mass jet
  head'' (see equations \ref{sol_xcm_t} and \ref{solution_rhoc}).
  The distance from the source $x_a$ corresponds either to the bow shock or
  jet shock positions (obtained from
  the numerical simulations), or to the position of the ``ram-pressure
  balance jet head'' (see equations \ref{xwsolm} and \ref{solution_ws_time}).
  Figure~\ref{fig:difverl} shows the relative position
  differences (with respect to the center of mass position for the jet head)
  for all of our computed models. The first thing to note (see the
  red lines in the two right panels of Figure 4) is that
  the relative position difference between the center of mass and
  ram pressure balance solutions does not strongly depend on time
  for the constant ejection density models (a result that can be
  seen by comparing equations \ref{solution_rhoc} and \ref{xxxws}).
  This result does not strictly hold for the constant mass loss rate models
  (red lines in the left panels of Figure 4).
  
  For the constant mass loss rate models we see that for $t>550$~yr
  the bow and jet shock positions remain below the center of mass
  working surface position, with negative $\Delta x$ values (see
  the left panels of Figure 4). For the constant ejection density models (right panels
  of Figure 4) we see an initial convergence to the center of
  mass solution of the bow and jet shock positions, and for
  $t>1300$~yr we see progressively larger, positive values for
  $\Delta x/x_{cm}$ (starting to approach the ram-pressure balance
  working surface position).

\section{Discussion}

We have presented a first attempt to compare the analytic
ram-pressure balance (Raga \& Kofman 1992) and center of
mass (Cant\'o et al. 2000) analytic formalisms (for obtaining the motions
of working surfaces in variable ejection jets) with axisymmetric
numerical simulations. To this effect, we choose the relatively
simple problem of the leading head of a jet produced with an ejection velocity that increases linearly with time.
We explore the cases of a time-independent mass loss rate
(in which the ejection density scales as the inverse of the
ejection velocity) and of a constant ejection density. From the simulations we compute
the on-axis positions of the jet shock and bow shock, and compare them with the predictions of the
analytic models (in which the jet and bow shock are assumed
to be spatially coincident).

For three of our four simulations (models $n_{100}$, $n_{5000}$ and
${\dot m}_{100}$), a good agreement is found
between the numerical simulations and the analytic models.
In these three models the jet and bow shock positions initially
follow the center of mass analytic solution, and (for two of these
models) at later
times deviate towards the faster moving, ram-pressure balance
analytic model. This result is completely satisfying, as one
would expect that at early times the shocked jet and environment material
will mostly remain within the working surface (as assumed
in the center of mass formalism), and that a substantial
sideways leakage of material (as assumed in the ram-pressure balance
model) should occur at later times. This is discussed in more
detail below.

It is clear that right after a working surface is formed,
the two associated shocks are very close to each other (with a separation
$d\ll r_j$, where $r_j$ is the local jet beam radius).
Because the surface $S\approx 2\pi r_jl$ through which the
gas exits the working surface is then small ($S\ll \pi r_j^2$), most
of the mass going through the shocks remains within the working surface.
At later times,
  the separation between the two shocks increases, and part of the
  mass processed by the working surface shocks is ejected
  sideways into a jet cocoon. For a jet with
    a time-indepenent ejection velocity, the separation $d$ between the two
    shocks attains a steady, maximum value determined by the balance between
    the incoming and outflowing mass rates. Such a balance between
    inflow and outflow from the working surface is
    also approximately attained in the case of a variable velocity
    jet, but the resulting shock separation (and also mass within
    the working surface) still has a (slower) time-dependence.

  It is possible to obtain
  analytic estimates of the separation between the two shocks
  (and therefore the mass of the gas within the working surface)
  with different degrees of complexity (and accuracy), following
  the approaches developed for determining the shock standoff
  distance in blunt body flows (see, e.g., the book of
  Hayes \& Probstein 2003 and Cant\'o \& Raga 1998). Here,
  we present a very simple estimate of the mass within a
  working surface.

  In the stationary state (i.e., after the initial regime of growing
  sepration between the two working surface shocks), the mass
  $M_{ws}$ within the two working surface shocks is:
  \begin{equation}
    M_{ws}\approx {\dot M}_{in}t_s\,,
    \label{mws}
  \end{equation}
  where ${\dot M}_{in}$ is the mass being fed through
  the two shocks and $t_s$ is the timescale for sideways
  leakage of the material. Assuming that the two shocks are
  highly radiative (with a cooling distance\footnote{The cooling distance is usually defined as the distance from a plane-parallel shock to the point where the temperature has dropped to a certain value (see for example Hartigan et al., 1987).} to $\sim 10^3$~K
  much smaller than the jet radius $r_j$), this leakage will
  occur at the post-cooling sound speed
  $c_0=\sqrt{5kT/3\mu m_H}\approx 3$~km~s$^{-1}$ (for $T=10^3$~K and
  $\mu=1.3$),
  and we then have
  \begin{equation}
    t_s= \frac{r_j}{2c_0}\,.
    \label{ts}
  \end{equation}
  Also, the total mass that has been fed into the working surface
  is
  \begin{equation}
    M_{in}\approx {\dot M}_{in}t_d\,,
    \label{min}
  \end{equation}
  where $t_d=x_{ws}/v_{ws}$ (with $x_{ws}$ being the position and
  $v_{ws}$ the velocity of the working surface) is the dynamical
  timescale.

  Combining equations (\ref{mws}-\ref{min}) we obtain\textbf{:}
  \begin{equation}
    \frac{M_{ws}}{M_{in}}\leq {\rm min}
      \left[1,\frac{r_j}{2x_{ws}}\frac{v_{ws}}{c_0}\right]\,.
      \label{mrat}
  \end{equation}
  In this equation, we have considered that the
  $M_{ws}/M_{in}\leq 1$ condition has to be satisfied
  (with $M_{ws}=M_{in}$ for the initial, mass conserving
  regime, and $M_{ws}<M_{in}$ when the sideways ejection from
  the working surface becomes important), while our analytic
  estimate can give values $>1$ because of its implicit assumption
  of a balance between the mass rates into and out of the working
  surface (which is not satisfied in the early evolution of the
  working surface, as discussed above). Clearly, for the
  case of an accelerating working surface (as the one that we
  obtain from our variable ejection models), the derivation of
  equation (\ref{mrat}) should be done evaluating the appropriate
  integrals (see section 4 of the paper of Lim et al. 2002),
  but the simple derivation shown above still gives
  the correct scaling properties.

  We now consider that the transition between the mass
    conserving and efficient mass loosing working surface
    regimes (which can be described with the ``center of mass''
    and ``ram-pressure balance'' equations of motion, respectively)
    occurs when $M_{ws}/M_{in}\sim 1/2$ (i.e., that the
    working surface has lost half of the processed jet material).
    Therefore, the position $x_t$ of the working surface
    at which the transition between the center of mass and ram
    pressure balance regimes occurs, can be estimated setting
    $M_{ws}/M_{in}=1/2$ in equation (\ref{mrat}), which gives:
  \begin{equation}
    x_t=\frac{v_{ws}}{c_0}r_j\,.
    \label{xtran}
  \end{equation}
  Setting $c_0=3$~km~s$^{-1}$ (see above), $r_j=300$~au (see
  section 3) and a velocity $v_{ws}\approx 200$~km~s$^{-1}$,
  we then obtain $x_t\approx 3\times 10^{17}$~cm, This is the correct
  order of magnitude for the location of the
  transition between the ``mass conserving''
  center of mass and the ``efficient mass loosing'' ram-pressure balance
  regimes for the working surface found in our
    $n_{100}$ and $n_{5000}$ numerical simulations (see Figure 3).

To summarize, we find that numerical simulations of a jet with
a linearly accelerating ejection velocity vs. time produce a
leading head which at early times can be analytically
modeled with the center of mass formalism, and (for appropriate flow
parameters) at later
evolutionary times approaches the ram-pressure balance solutions
(the distance from the source at which the
transition between these two regimes occurs can be estimated
with equation (\ref{xtran})).
This result holds provided that during its time-evolution the
jet does not become under dense/under pressured, in which case
strong departures from the analytic solutions are found.

\section{Conclusions}

This paper presents a first attempt to compare analytic solutions
based on both the ``ram-pressure balance'' and ``center of mass''
formalisms with hydrodynamical (axisymmetric) simulations. For
this comparison, we have studied the case of a linearly increasing
ejection velocity as a function of time (and two different forms
for the ejection density, see \S 2).

This simple ejection variability could apply to the early evolution
of an HH outflow, and is relevant for the problem of entrainment
of environmental molecular material into the jet head (see Lim et al. 2002)
and the potential destruction of molecules by shocks, particularly in the jet head. As a direct application to the observed, evolved HH
jets is not straightforward, the interest of our work is primarily
theoretical.

While full analytic solutions for different forms of the ejection variability 
have been found in the past with the ``center of mass'' formalism
(see Cant\'o et al, 2000 and Cant\'o \& Raga 2003), no solutions
(except the trivial one for the constant ejection case) have been
previously found with the ``ram-pressure balance'' equation of
motion for the working surfaces. We were surprised to find that
the ``linearly accelerating'' ejection velocity case does lead
to a full, analytic ram-pressure balance solution for the case of
a constant ejection density (though we managed to find only an
approximate analytic solution for the constant mass loss rate
case). This result \textbf{implies} that ram-pressure balance
analytic solutions might also exist for other ejection velocity/density
variability combinations.

Comparing the center of mass and the ram-pressure balance solutions
with axisymmetric numerical simulations (with the same parameters)
we find that for the case of over-dense jets (with
higher densities in the jet side of the leading working surface)
the numerical results initially follow the center of mass
solution, and at later times approach the ram-pressure balance
analytic solution. This result shows a satisfying consistency
between the analytic approaches and the numerical solutions with
the full (axisymmetric) hydrodynamic equations. Not entirely
unexpectedly, we do not
find a good agreement between the analytic and numerical
results for cases in which the jet density drops to values
lower than the environmental density, as the jet beam then develops
internal shocks (in the numerical solutions) that affect the motion
of the leading working surface.

Clearly, it would be interesting to extend this comparison between
numerical simulations and analytic solutions of variable jets
to the case of jets with ``internal working surfaces'' within
their beams. This problem, which would clearly have much more direct
applications to observed HH jets, is left for a future study.

\vspace{0.25cm}
ARa and JC acknowledge support from the DGAPA-UNAM grant IG100218.
We thank an anonymous referee for helpful comments which lead
(among other things) to many of the issues discussed in section 5.

\vfill\eject
$\,$
\vfill\eject
\begin{appendix}
\section{Appendix}

\begin{figure}
 \centering
  \includegraphics[scale=0.535]{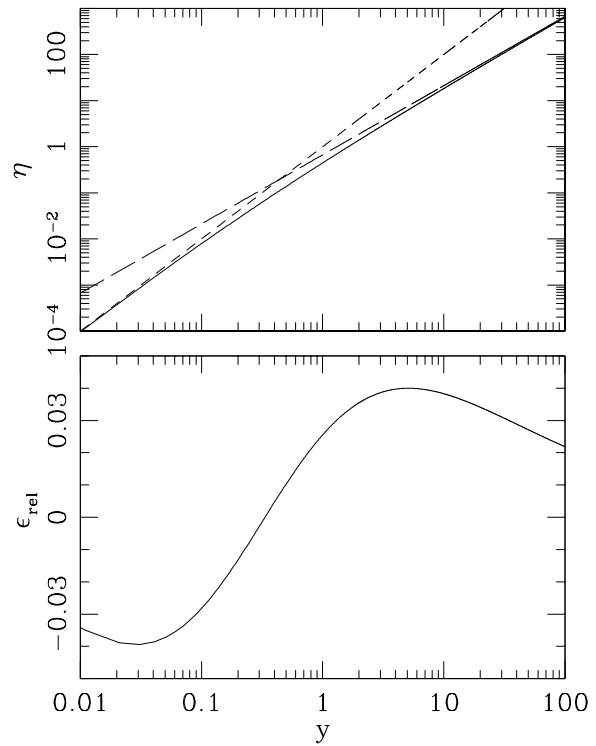}
  \caption{\footnotesize{Top frame: the exact (i.e., numerical) solution of equation
    (\ref{dimensionless_eqmov}) (solid curve), the ``near'' (short
    dashes) and the ``far field'' (long dahes) analytic solutions given
    by equation (\ref{near_far}). Bottom frame: relative deviation of the
    approximate solution of equation (\ref{average}) with respect to
    the ``exact'' (numerical) solution of equation (\ref{dimensionless_eqmov}).}} 
    \label{rpb}
\end{figure}

In \S 2.1 we established that for a constant mass loss rate
the ram-pressure balance assumption leads to equation
(\ref{ctemasslosseqqdiff}). In terms of dimensionless variables
(see equation \ref{dimensionless_var}) the equation of motion takes
the form presented in (\ref{dimensionless_eqmov}).
By integrating it numerically we obtain the $\eta(y)$ dependency
shown the top frame of Figure \ref{rpb}.

This figure also shows the``near'' and ``far field'' solutions
    \begin{equation}
    \eta_{n}(y) = y^2 \;,\;\;\;\;\;
    \eta_{f}(y) = \frac{2}{3}y^{3/2},
    \label{near_far}
    \end{equation}
which correspond to the $y \rightarrow 0$ (near) and
$y \gg 1$ (far) limits, respectively. These two solutions
can be obtained straightforwardly  taking the appropriate
limits in equation (\ref{dimensionless_eqmov}).

Since the solutions in (\ref{near_far}) represent two regimes with
analytical integrals, a non-linear average between this approximations
is computed:
    \begin{equation}
        \eta^{(a)} = \left(\eta_n^{-5/4} + \eta_f^{-5/4} \right)^{-4/5}.
        \label{average}
    \end{equation}
to obtain an approximation to the full $\eta(y)$ solution.
    These average has a relative error
    $\epsilon_{rel} = \eta^{(a)}/\eta -1<0.04$ with respect to the
    ``exact" solution $\eta(y)$ (obtained by
      numerical integration of equation \ref{dimensionless_eqmov}), and
    is therefore a reasonable analytical approximation of the solution. 
    The dependency of $\epsilon_{rel}$ on $y$ is shown in the bottom panel
    of Figure \ref{rpb}.
    
     Another approximation to the numerical $\eta(y)$ solution is
    \begin{equation}
        \eta^{(b)} = \frac{2y}{9} \left(-1+\sqrt{1+9y} \right),
    \end{equation}
    which coincides with the exact solution in the $y\rightarrow0$
    and $y\rightarrow\infty$ limits, and has a maximum relative
    deviation of $\approx 0.09$. This approximation has a functional
    form similar to the solution for the constant
    mass loss rate case obtained with the ``center of mass formalism''
    (see equation \ref{mdotcte_solution}).

\end{appendix}

\end{document}